\documentclass[aps,twocolumn,graphics,floatfix,tightenlines]{revtex4-2}
\usepackage{graphics}
\usepackage{epstopdf}
\usepackage{epsfig}
\usepackage{graphicx}
\usepackage{epsf,epic}
\usepackage{color}
\usepackage{subfig}
\usepackage{amsmath}
\usepackage{booktabs}
\usepackage{multirow}
\usepackage{physics}
\usepackage{hyperref}
\usepackage{amsfonts}
\usepackage{wrapfig}
\usepackage{breqn}
\usepackage{pstricks}
\usepackage[utf8x]{inputenc}
\usepackage{multirow}
\usepackage{fancyref}
\usepackage{pst-node}
\usepackage{float}
\usepackage{bm}
\usepackage{dcolumn}
\newcommand{\etal}{\textit{et al.\ }}

\makeatletter
\usepackage{etoolbox} 
\appto{\appendix}{%
	\@ifstar{\def\theequation@prefix{A.}}%
	{}%
}
\preto\maketitle{%
  \begingroup\lccode`~=`,
  \lowercase{\endgroup
  \let\saved@breqn@active@comma~
  \let~}\active@comma 
}
\appto\maketitle{%
  \begingroup\lccode`~=`,
  \lowercase{\endgroup
  \let~}\saved@breqn@active@comma 
}
\makeatother
\begin{document}
\title{Optical response and band structure of LiCoO$_2$ including electron-hole
  interaction effects}
\author{Santosh Kumar Radha}
\author{Walter R. L. Lambrecht}\email{walter.lambrecht@case.edu}
\affiliation{Department of Physics, Case Western Reserve University, 10900 Euclid Avenue, Cleveland, OH-44106-7079}
\author{Brian Cunningham}
\author{Myrta Gr\"uning}
\altaffiliation[Also at ]{European Theoretical Spectroscopy Facility (ETSF)}
\affiliation{School of Mathematics and Physics, Queen’s University Belfast, Belfast BT7 1NN, Northern Ireland, United Kingdom}
\author{Dimitar Pashov}
\author{Mark van Schilfgaarde}
\altaffiliation[Also at ]{National Renewable Energy Laboratory, Golden, CO 80401, USA}
\affiliation{Department of Physics, King's College London, London WC2R 2LS, United Kingdom}
\begin{abstract}
  The optical response functions and band structures of LiCoO$_2$ are studied
  at different levels of approximation, from density functional theory (DFT)
  in the generalized gradient approximation (GGA)  
  to quasiparticle self-consistent QS$GW$ (with $G$ for Green's function and $W$ for screened Coulomb interaction)
  without and with ladder diagrams (QS$G\hat W$) and
  the Bethe Salpeter Equation (BSE) approach.  The QS$GW$ method is found to
  strongly overestimate the band gap and electron-hole or excitonic effects
  are found to be important. They lower the quasiparticle gap by only about 11~\% but the lowest energy peaks in absorption are
  found to be excitonic in nature. The contributions from different
  band to band transitions  and the relation of excitons
  to band-to-band transitions are analyzed. The excitons
  are found to be strongly localized.  A comparison to  experimental data  is presented.  
\end{abstract}
\maketitle
\section{Introduction}
LiCoO$_2$ is a well known material utilized in Li-ion batteries.\cite{Mizushima80,Miyoshi18,Iwaya13} In spite of the extensive literature on this material,
its fundamental electronic structure and optical properties are not
yet fully understood. Qualitatively, its electronic structure is understood
to be a normal band insulator with a low spin configuration in terms of
the $d$-band filling.\cite{Czyzyk92,vanElp}
The structure with $R\bar{3}m$ space group consists
of layers of edge-sharing CoO$_6$ octahedra with a triangular Co lattice, 
with intercalated Li. One may also view it as a layered ordered arrangement
of Li and Co ions in a close packed oxygen lattice with both Li and Co
octahedrally coordinated. In contrast, another form of LiCoO$_2$ with
disordered spinel structure has both octahedral and tetrahedrally coordinated
cations. In the $R\bar{3}m$ structure, 
the Li donates its electron to the Co-O layer,
thus leading nominally to a Co$^{3+}$ ion with $d^6$ configuration in which
the lower Co-$d$ $t_{2g}$ orbital derived bands are filled and the $e_g$ bands 
are empty.\cite{Aydinol} Much attention has been paid to the effects of
delithiation and ordering of Li vacancies
in Li$_x$CoO$_2$,\cite{Wolverton,Marianetti04,Iwaya13,Miyoshi18}
including the  $x=0$ end member CoO$_2$.\cite{Seguin99}

However,  a quantitative understanding of optical absorption in relation
to the band structure is missing. On the experimental side,
a lot of confusion seems to arise from
the variations in Li content in Li$_x$CoO$_2$ and dependence on growth
methods, temperature dependent variations in the structure of LiCoO$_2$
and the deviations from single crystal behavior. Different
methods of determining the optical absorption, such as Tauc plots
vs. reflection or transmission measurements also tend to give different
results. On the theory side, band structures in the  local density
approximation\cite{Czyzyk92,Aydinol}  of density functional theory seemed to give qualitative agreement with the optical absorption onset near 1.5 eV 
 but, as is a  main focus of this paper,
the higher accuracy $GW$ many-body-theory results give a much higher gap. 
A detailed comparison of our calculations with specific measurements is
postponed till Sec. \ref{sec:expt}.
The previous theory assignments of the optical features were based only
on peak positions but not on actual calculations of the optical response. 

In this paper we calculate and analyze the optical response functions
based on band structure calculations  at different levels of theory.
We start by comparing density functional calculations in the generalized
gradient approximation (GGA) to quasiparticle self-consistent
$GW$ calculations. Next, we calculate the
imaginary part of the dielectric function $\varepsilon_2(\omega)$ first
in the long-wave length approximation from a summation over the interband
transitions and analyze the contributions from different bands to
the main peaks based on the QS$GW$ bands.
We then calculate $\varepsilon_2(\omega)$ including
local field effects in the random phase approximation (RPA) and using the
Bethe-Salpeter Equation (BSE) method including thereby electron-hole interaction effects. The electron-hole interaction effects or ladder diagrams
in the calculation of $W$,
the screened Coulomb interaction are also included here at finite
wavevector $q$ and we call this band structure the QS$G\hat W$  band
structure because it includes vertex ($\Gamma$) corrections to $W$. 

We find that the QS$GW$ band gap is significantly higher ($\sim$4 eV),
than the GGA  gap, which is not unusual, but  also larger
than the experimental values.
While QS$G\hat W$  reduces the $\Sigma$ correction by about 11 \%, the gap
is still significantly higher than experiment. 
We show that the lowest peaks in absorption are in fact
excitonic in nature and significantly lower the optical gap $E_g^{opt}$
compared to the fundamental one-particle gap, which is defined as the
difference between the ionization potential and the electron affinity,
 $E_g^{qp}=I-A$. 
The electron affinity is defined as the 
lowest energy for adding an electron ($A=E(N)-E(N+1)$)
ad the ionization potential as the energy for extracting
an electron ($I=E(N-1)-E(N)$).

  \begin{figure*}[ht]
  \includegraphics[width=\linewidth]{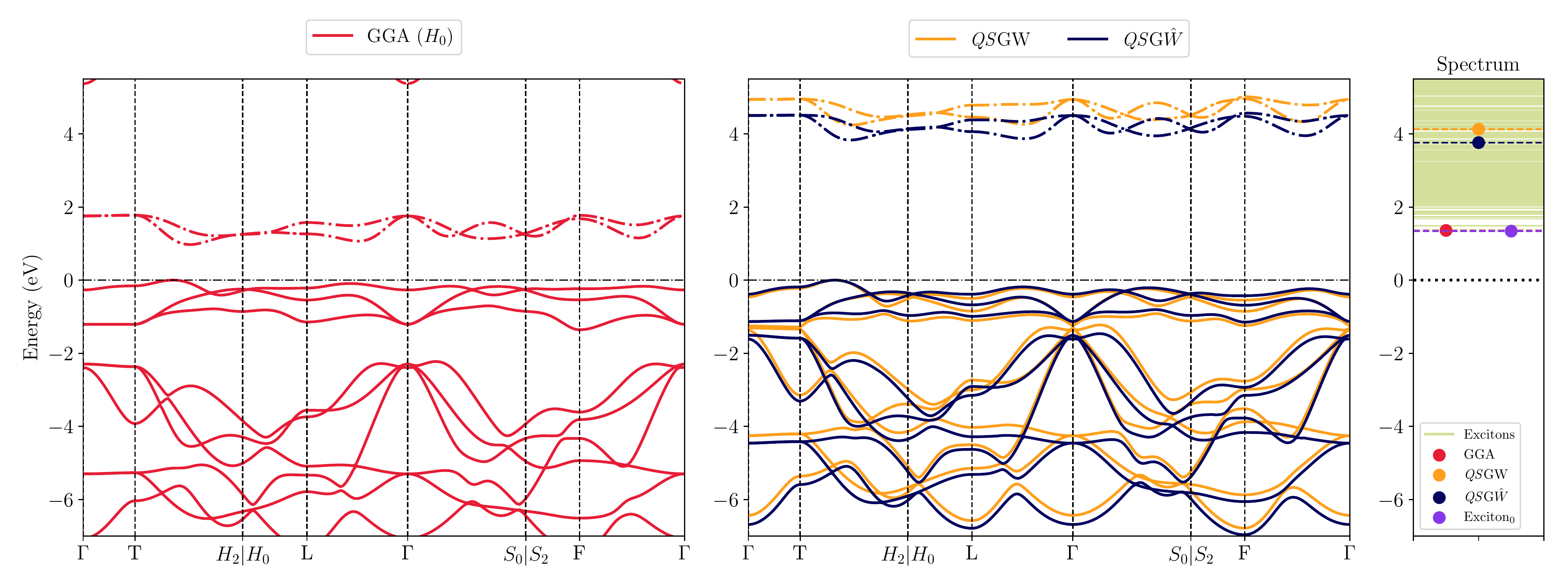} 
  \caption{Band structure of $R\bar{3}m$ LiCoO$_2$ in
    GGA (left), and  QS$GW$, QS$G\hat W$  (middle) ; 
    The Brillouin zone high symmetry point labeling follows Refs.\cite{Aroyo14,Hinuma17}.
    The rightmost figure shows the exciton spectrum of eigenvalues of the two particle BSE compared with
    the band gaps at different levels of theory. The $\hat W$ is used here for the exciton levels.\label{figldagw}} 
  \end{figure*}
\section{Computational Methods}
All the calculations done here make use of the full-potential linearized muffin-tin orbital method (FP-LMTO) \cite{Kotani10,Methfessel}
as implemented in the Questaal package.\cite{questaalpaper,questaal} This is an all-electron method without shape approximations
to the potentials. In the density functional calculations, we use the
generalized gradient approximation (GGA)
in the Perdew-Burke-Ernzerhof (PBE)\cite{PBE} parameterization.
The quasiparticle self-consistent $GW$ method (QS$GW$)
is described in Ref. \onlinecite{Kotani07}. It is based on Hedin's many-body-perturbation theoretical method in which the dynamical
self-energy is given schematically
as $\Sigma(\omega)=G(\omega)\otimes W(\omega^\prime)$ with $\otimes$ meaning
convolution in energy and {\bf k}-space.
Here $G(\omega)$ is the one-particle Green's function corresponding
to the starting independent particle $H^0$ Hamiltonian
(usually the  local density approximation (LDA) or GGA at first),
and $W(\omega)$ is 
the screened Coulomb interaction, $W(\omega)=[1-vP(\omega)]^{-1}v$
in which $v$ is the bare Coulomb interaction and $P(\omega)$ the
polarization propagator, which can also be obtained from the Green's function
$G$  or directly from the eigenvalues $\epsilon_i$
and eigenstates $|\psi_i\rangle$
of the $H^0$ Hamiltonian. Here the index $i$ is a short hand for band number
$n$ and ${\bf k}$-point in the Brillouin zone. 
From this self-energy, a non-local but energy independent exchange-correlation
potential 
$\tilde\Sigma_{ij}=\frac{1}{2}\mathrm{Re}\{\Sigma_{ij}(\epsilon_i)+\Sigma_{ij}(\epsilon_j)\}$
is extracted in the basis of the eigenstates  of the $H^0$
Hamiltonian. This new exchange-correlation potential now replaces the
one in $H^0$ and the procedure is iterated till convergence, at which point
the eigenvalues $\epsilon_i$ of $H^0$ become identical to the quasiparticle
energies $E_i$ which have the meaning of one-electron extraction and addition
energies to the many-electron system.

Note that in the above procedure the RPA
is usually used in calculating $W$. This tends to underestimate the screening.
A significant improvement consists in including electron-hole interactions
or ladder diagrams in the calculation of $W$ via the BSE as described in
Cunningham \etal\cite{Cunningham18,Cunningham21}. We'll denote this new $W$
by $\hat W$. Please note that it  plays a similar role to $\tilde W$
in other approaches which use a time-dependent DFT kernel $f_{xc}$ to
correct $W$.\cite{Shishkin07,ChenPasquarello15,Tal21}
Our approach is equivalent to including a specific approximation for the vertex in
the calculation of $P$ in the context of the Hedin equations, namely $\delta \Sigma(12)/\delta G(34)=iW(12)\delta(13)\delta(24)$ is evaluated
from $\Sigma=iGW$ assuming that $\delta W/\delta G\approx0$. Here the numbers represent real space, spin and time variables
$1\equiv({\bf r}_1,\sigma_1,t_1)$.
The vertex then becomes
\begin{equation}
  \Gamma(123)=\delta(12)\delta(13)+i\int d(67) W(12)G(16)G(72)\Gamma(673)
\end{equation}
and introducing the 4-point $P^{RPA}(1234)=-iG(13)G(42)$ allows us to cast this in terms of a BSE approach,
in which $P(12)=P(1122)$ is obtained from
\begin{eqnarray}
  P(12)&=&P_{RPA}(12)-\nonumber \\
  &&\int P_{RPA}(1134)W(34;\omega=0)P(3422) d(34).\nonumber \label{eq:Pladder}\\
\end{eqnarray}
As usual, we here make a static approximation for $W$ for the inclusion of the electron-hole effects.  This has mainly
been validated by its success in describing optical response. Going beyond it would significantly increase the
complexity of the method.\cite{Marini03} Marini and Del Sole\cite{Marini03} showed that including the exciton dynamics
partially undoes the renormalizaton effects of the quasiparticle self-energy. However, this may be different in our
QS$GW$ approach which relies on error-cancellations of such renormalization effects.

At present, we do not include a
corresponding vertex correction in the calculation of the self-energy.
Gr\"uneis \etal \cite{Gruneis14} found first-order vertex corrections in the self-energy to improve mainly
the absolute ionization potentials while only moderately affecting band gaps
for various weakly correlated semiconductors. For small molecules,
Maggio \etal \cite{Maggio17} found  that when applying $G_0W_0\Gamma$ to a Hartree-Fock starting point, including the vertex in $W$
improved ionization potentials while including the vertex also for the self-energy worsened results. 
While vertex corrections in $\Sigma$ were found to be important in the
context of self-consistent implementation of the Hedin equations by Kutepov\cite{Kutepov16,Kutepov17}, our quasiparticle
self-consistent approach has a different aim, namely to provide the best one-particle starting point for $G$ and the
best possible screening in $W$ to provide accurate quasiparticle energies within the $GW$ approximation.
As we showed in our introduction of the QS\emph{GW} approximation, quasiparticalization of $G$ greatly
  reduces errors introduced by omitting $\Gamma$ in $\Sigma$ (See Appendix A, reference \onlinecite{Kotani07}.  Calling $G^{0}$ the quasiparticlized $G$, we can write $G^{}=ZG^{0}+{\bar{G}}$, where ${\bar{G}}$ is the
  incoherent part. On the other hand, in the $\omega{\rightarrow}0,\, \mathbf{q}{\rightarrow}0$ limit there is a Ward
  identiy $\Gamma{\rightarrow}1/Z$.  This $Z$ factor approximately cancels the one in $GW$ with the replacement $G{\rightarrow}G^{0}$.
While it would be of interest to study the effect of vertex corrections in $\Sigma=iGW\Gamma$, as the equations are
somewhat imbalanced by including only the vertex in $W$, it is beyond the scope of the present paper.

The optical response is given in terms of the macroscopic dielectric function,
in particular its imaginary part $\varepsilon_2(\omega)$, from
which the real part $\varepsilon_1(\omega)$ can be obtained by Kramers-Kronig
transformation and from it all other relevant optical functions, such
as the complex index of refraction, absorption coefficient and reflectivity. 
A first way to calculate this is through the Adler-Wiser equation
in the independent particle, long-wave length limit.
\begin{eqnarray}
\varepsilon_2(\omega)&=&\frac{8\pi^2e^2}{\Omega \omega^2}\sum_n\sum_{n'}\sum_{{\bf k}\in BZ}f_{n{\bf k}}(1-f_{n'{\bf k}}) \nonumber \\
&&|\langle \psi_{n{\bf k}}|[H,{\bf r}]|\psi_{n'{\bf k}}\rangle|^2\delta(\omega-\epsilon_{n'{\bf k}}+\epsilon_{n{\bf k}}). \label{eqAdlerWiser}
\end{eqnarray}
The optical matrix elements here are the velocity matrix elements,
$\dot{\bf r}=(i/\hbar)[H,{\bf r}]$, which for a local potential
can be written in terms of the momentum matrix elements
${\bf v}={\bf p}/m$. So, when using the LDA or GGA band structure and
eigenstates, this is correct but when using the $GW$ band structure
as input, one needs to renormalize the matrix elements.
One way to do this, proposed by Levine and Allan \cite{Levine89} in
the context of a scissor-operator corrections, 
consists in rescaling the matrix elements 
by a factor $(\epsilon_{n'{\bf k}}-\epsilon_{n{\bf k}})/(\epsilon^{LDA}_{n'{\bf k}}-\epsilon^{LDA}_{n{\bf k}})$.
Alternatively $d\tilde\Sigma/dk$ terms need to be explicitly included.  We follow the latter. 

A second, more accurate formulation is to calculate
\begin{equation}
  \varepsilon_M(\omega)=\lim_{{\bf q}\rightarrow0}\frac{1}{\varepsilon^{-1}_{{\bf G}=0,{\bf G}^\prime=0}({\bf q},\omega)} \label{eqRPA}
\end{equation} 
in a basis set of plane waves. This formulation includes local-field
effects, whereas the above Adler-Wiser equation does not.
Within this formulation, one may now either approximate
$\varepsilon_{{\bf G},{\bf G}^\prime}^{-1}({\bf q},\omega)$ by including
the Coulomb interaction in $\varepsilon^{-1}=(1-vP)^{-1}$, but 
neglecting elecron-hole interactions (carried by $W$ in Eq. \ref{eq:kernel})
called the RPA,  
or by using the BSE including electron-hole interactions
as described in Cunningham \etal\cite{Cunningham18}.
This step is formulated in terms of four-particle polarization operators
and more specifically, one calculates the
modified response function\cite{Hanke78,Onida02}
\begin{equation}
  \bar{P}(1234)=P^0(1234)+\int d(5678)P^0(1256)K(5678)\bar{P}(7834) \label{eq:BSE1}
\end{equation}
with the  kernel,
\begin{equation}
  K(1234)=\delta(12)(34)\bar{v}-\delta(13)\delta(24)W(12). \label{eq:kernel}
\end{equation}
Again, we may also here replace
$W$ by $\hat W$ for an even better approximation and we use a static approximation $W(\omega=0)$
in solving the BSE equation \ref{eq:BSE1}.\cite{Cunningham21}
Finally, note that
$\bar{v}_{{\bf G}}({\bf q})=4\pi/|{\bf q}+{\bf G}|^2$ if ${\bf G}\ne0$
and zero otherwise. 
The macroscopic dielectric function is then given by
\begin{equation}
  \varepsilon_M(\omega)=1-\lim_{{\bf q}\rightarrow0}v_{{\bf G}=0}({\bf q})\bar{P}_{{\bf G}={\bf G}^\prime=0}({\bf q},\omega) \label{eqepsmac}
\end{equation}
Essentially, this approach amounts to analytically finding the relevant matrix element ${\bf G}={\bf G}'=0$ of the inverse
of the dielectric matrix by a block matrix inversion approach as explained in appendix B of Ref. \onlinecite{Onida02}.
In practice the Dyson equation for  $\bar{P}$ is solved in the basis
set of single-particle eigenfunctions as described in Cunningham \etal\cite{Cunningham18}.
Eventually, the bare and screened Coulomb matrices in the kernel
are first written in terms of the mixed product basis set,
($|M^{{\bf q}}_I\rangle$) used also
in the $GW$ implementation and then converted to the single-particle basis set
to derive an effective two-particle Hamiltonian which is then diagonalized
to obtain the final forms of the macroscopic dielectric function $\varepsilon_M(\omega)$ including the optical matrix elements,
(see Eqs. 23-25 in Ref. \onlinecite{Cunningham18}). 
The optical matrix elements of the velocity operator are obtained
including the momentum derivative of the self-energy.\cite{Cunningham18} 
The use of the mixed product/interstitial plane wave auxiliary basis was designed to represent the polarization $P$
more efficiently than is possible with traditional plane-wave basis sets.  In the latter case the polarizability is
obtained by the Adler-Wiser construction, which involves terms of a sum over all empty eigenstates. It is in principle
exact for the RPA polarizability, but it is known to converge slowly and alternative schemes have been proposed.\cite{Loos2020,Berger2010,Betzinger2015}.
The situation is quite different for the basis we use: quasiparticle levels
converge very quickly with the rank of the basis set, as was shown in some detail for a number of semiconductors.\cite{vanSchilfgaarde06}
This rapid convergence is slightly misleading, however, because in the augmentation region the basis set does not change
except with the addition of local orbitals.  Our present scheme allows the use of one local orbital per $l$-channel, but it has been
shown \cite{Betzinger13} that for full convergence, either more than
one is required or that the partial waves are generalized
to be frequency dependent.\cite{Betzinger2015}
This error is fairly small: in the case of ZnO, which is a worst
case scenario, the discrepancy between a fully converged result
\cite{Friedrich11}  and the present method
in the $G_0W_0$ approximation \cite{vanSchilfgaarde06} is approximately 0.2 eV.
High-energy local orbitals were also shown to be important in
Ref.\onlinecite{Jianghong16}.

Adding ladder diagrams in the calculation of $W$ proceeds similarly
by solving the Dyson-like Eq.(\ref{eq:Pladder}).\cite{Cunningham21}
This is done in the ``transition space'' by expanding the 4-point
quantities in one-particle eigenfunctions.
We then solve a BSE equation for the two-particle Hamiltonian:
\begin{eqnarray}
  H_{n_1n_2n_3n_4}&=&(\epsilon_{n_2}-\epsilon_{n_1})\delta_{n_1,n_3}\delta_{n_2,n_4}\nonumber \\ &&+
  (f_{n_4}-f_{n_3})W_{n_1n_2n_3n_4}(\omega=0)
\end{eqnarray}
where $f_n$ are Fermi functions. 
We work within the Tamm-Dankoff approximation (TDA), and using a static
$W(\omega=0)$,  which means that
$n_1=v$, $n_2=c$, $n_3=v'$, $n_4=c'$ are restricted to be
valence and conduction band states. The TDA has been found to be adequate for calculation of optical
absorption, in particular when combined with the modified response function approach by Hanke \cite{Hanke78} as shown by
explicitly going beyond it in Ref. \onlinecite{Sander15}. Its validity for finite ${\bf q}$ is less well established but
going beyond the TDA significantly increases the computational effort by requiring a non-Hermitian and double sized matrix to
be inverted. While including here finite $q$ electron-hole interactions, we expect their major effect on the response to
occur for the long-wavelength limit. 
After diagonalizing this two-particle
Hamiltonian, 
\begin{equation}
  H_{vc,v'c'}A^\lambda_{v',c'}=E_\lambda A^\lambda_{v,c}
\end{equation}
The 4-particle polarization is contracted back to a two particle one
and re-expressed in the mixed-product basis set. We should also
keep in mind that each valence and conduction state here are associated
with a different ${\bf k}$-point in the Brillouin zone. We then obtain
\begin{eqnarray}
  P_{I,J}({\bf q},\omega) &=&\sum_{v,c,v',c'}\sum_{{\bf k},{\bf k}'}
  \langle M_I^{{\bf q}}\psi_{v{\bf k}}|\psi_{c{\bf k}+{\bf q}}\rangle \nonumber\\
  &&\sum_\lambda\frac{A^\lambda_{v{\bf k},c{\bf k}+{\bf q}}\left[A^\lambda_{v'{\bf k}',c'{\bf k}'+{\bf q}}\right]^*}{E_\lambda-\omega+i\eta}\nonumber \\
  &&\langle \psi_{v'{\bf k}'}|\psi_{c'{\bf k}'+{\bf q}}M_J^{{\bf q}}\rangle
\end{eqnarray}
Finally, the new $W$ is obtained as
\begin{eqnarray}
  W_{I,J}({\bf q},\omega)&&=\sqrt{V_I({\bf q})}\left[ 1-\sqrt{V({\bf q})}P({\bf q},\omega)\sqrt{V({\bf q})}\right]^{-1}_{I,J}\nonumber \\\sqrt{V_J({\bf q})}
\end{eqnarray}
We work here in the product basis set which is rotated so that the
bare Coulomb interaction is diagonal and its square root can
be taken. 
  
 We note that obtaining $\hat W({\bf q},\omega)$ requires diagonalizing
the two particle Hamiltonian of the 
BSE equations for a mesh  of ${\bf q}$ points
and is therefore more demanding
than the final form of the macroscopic dielectric function (Eq. \ref{eqepsmac})
which
only requires diagonalizing the two-particle Hamiltonian
in the limit ${\bf q}\rightarrow0$. Also using the expression
for the macroscopic dielectric function in terms of the
modified response function $\bar{P}$ avoids having to invert the full
dielectric response matrix. 

Calculations are done in the full-potential linearized muffin-tin orbital method. Convergence parameters were chosen as follows:
basis set $spdf −spd$ spherical wave envelope functions plus augmented plane waves with a cut-off of 3 Ry, augmentation cutoff $l_{max} = 4$, {\bf k}-point mesh, $12\times12\times12$. In the $GW$ calculations the self energy $\Sigma$ is calculated on a f {\bf k}-mesh of $6\times6\times6$ points and
interpolated to the above finer mesh and the bands along symmetry lines using the real space representation of the LMTO basis set.
\begin{figure}
  \includegraphics[width=\linewidth]{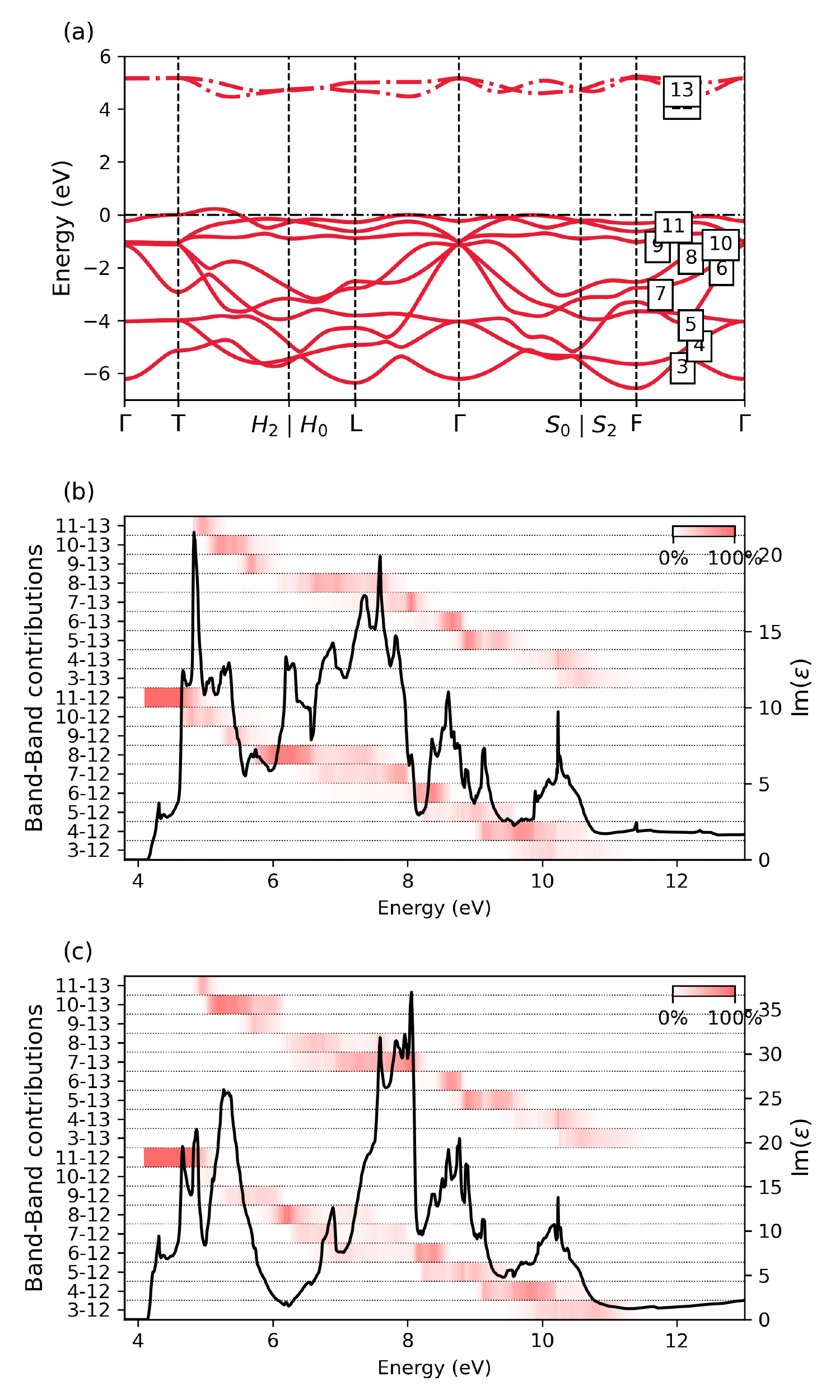}
  \caption{Interband optical response function $\varepsilon_2(\omega)$
    calculated in long-wavelength approximation (no local field or excitonic effects). 
    The  partial contribution from a given band pair to the spectrum are shown as a color scale for each band pair:  (a) band numbering (QS$G\hat W$)
    (b)  ${\bf E}\parallel {\bf c}$, (c) ${\bf E}\perp{\bf c}$,  \label{figepslmf}}
\end{figure}

\section{Results}
\subsection{Band structure}

  First, we compare the GGA, QS$GW$ and QS$G\hat W$  band structures
  of LiCoO$_2$ in the $R\bar{3}m$ structure in 
  Fig. \ref{figldagw}. The Brillouin zone labeling follows the notation of
the Bilbao Crystallography Server(\url{/https://www.cryst.ehu.es})
which is equivalent to  Hinuma \etal \cite{Hinuma17,seekpath}.
The band gap is found to be slightly indirect with VBM and 
CBM lying between $T$ and $H_2$ very close to each other but not exactly
at the same point. The lowest indirect gap is thus very close to
direct gaps at the VBM and CBM or the lowest direct gap. 
The gaps of interest and {\bf k}-location of band extrema
are summarized in Table \ref{tabgaps}. 
We can see that the QS$GW$ gap (4.125 eV) is significantly larger than the
experimental values mentioned in the introduction.
The GGA gap is somewhat smaller than the reported $t_{2g}-e_g$ gap of about 2 eV and smaller
than the value in Czy\'{z}yk \etal\cite{Czyzyk92} of 1.2 eV.
This is related to our use of the GGA  lattice constants. 
We note that $(E_g^{QSG\hat W}-E_g^{GGA})/(E_g^{QSGW}-E_g^{GGA})\approx0.89$.
Thus, adding the ladder diagrams to include the electron-hole effects on the screening of $W$
reduces the gap correction by about 11 \%, somewhat smaller than the often used  {\sl ad-hoc}
0.8$\Sigma$ correction  factor.

We may also see that the Co-$d$-$t_{2g}$ bands (between 0 and $-2$ eV in GGA)
have moved closer to the more O-$2p$ like deeper valence bands. We should
note that we have used the VBM as reference for both. So, what this really
indicates is that the Co-$d$-$t_{2g}$ bands shift more down by the $GW$
self-energy than the O-$2p$, which results from their more localized
character.
In fact, the top valence band in QS$G\hat W$ shifts down by about 1.33 eV
relative to GGA,  the conduction band shifts up by about 1.53 eV and
the fourth valence band counting down from the top, which is the top of
the O-$2p$ like bands shifts down by only 0.42 eV at $\Gamma$.

Finally, we note that convergence of the self-energy with $\hat W$ is important. We found that the band gap converges faster to the final result if we
apply the ladder diagrams in $\hat W$ from the start rather than first
doing a QS$GW$ calculation and then adding the ladder diagrams.
This is shown in Appendix\ref{sm:convergence}.
\begin{figure}[h]
  \includegraphics[width=\linewidth]{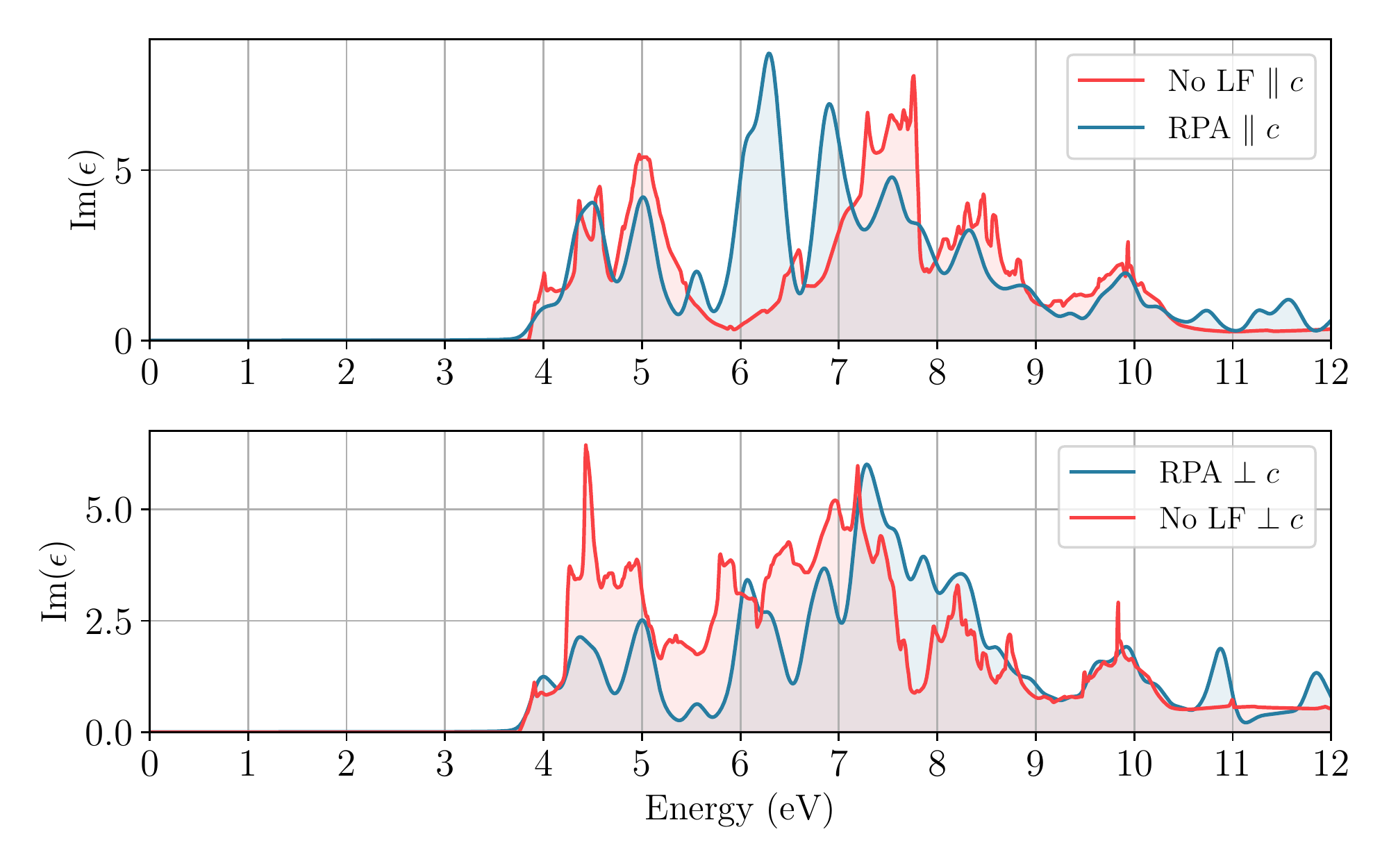}
  \caption{Macroscopic dielectric function $\varepsilon_2(\omega)$ for both polarizations comparing  RPA (Eq. \ref{eqRPA}) with the
    long-wavelength limit (Eq. \ref{eqAdlerWiser}) without local field effects.\label{fig:rpalmf}}
\end{figure}

\begin{table*}
  \caption{Band gaps of $R\bar{3}m$ LiCoO$_2$ in various approximations
    and {\bf k}-location of VBM and VBM.\label{tabgaps}}
\begin{ruledtabular}
  \begin{tabular}{llll}
\hline
             & GGA                 & QS$GW$                & QS$G\hat W$      \\
\hline
Smallest indirect gap (eV)    & 0.867               & 4.125               & 3.762              \\
Direct gap at $\Gamma$ (eV)            &     2.023                &    5.401             & 4.884       \\
$\textbf{ k}_{CBM}$  & (0.612,0.388,0.500) & (0.604,0.396,0.500)  & (0.604,0.396,0.500)  \\
$\textbf{ k}_{VBM}$  & (0.604,0.396,0.500) & (0.596,0.404,0.500) & (0.596,0.404,0.500)\\
\hline
  \end{tabular}
\end{ruledtabular}
\end{table*}

\subsection{Band to band analysis of optical response}
Next, we show the optical response function obtained within the Adler-Wiser form
using the QS$G\hat W$  bands and matrix elements  in \autoref{figepslmf}. 
Along with it, we show the corresponding band structure. The relevant bands
are numbered. The vertical axis in the $\varepsilon_2(\omega)$ figure
is divided in intervals corresponding to specific valence band to
conduction bands as numbered in the band figure and the horizontal
color bars  show their partial contribution to the spectrum on a
color scale shown on the right. 
For example, for ${\bf E}\perp{\bf c}$
the lowest energy peak, just above 4 eV
is almost 100 \% accounted for by the transition from the top valence band
(no. 11) to the lowest conduction band (12) while the peak just above 6 eV
is mostly accounted for by bands 8 to 12 transitions.  The peak at about 5 eV
has a large contribution from bands 10 to 13 transitions. 
We can also see that there is a significant anisotropy between the
${\bf E}\perp{\bf c}$ and ${\bf E}\parallel{\bf c}$ response.
Finally, we note that all peaks up to 10 eV are mainly explained by transitions to the lowest two
conduction bands from increasingly deeper valence bands. Transitions from the top of the valence band to
higher Co-$4s$ or Li-like band do not appear to  make a significant contribution in this range.
Details of the higher bands can be found in Ref. \onlinecite{Volkova21}
in the GGA. They lie above 6 eV in that case and thus even higher, above
10 eV in QS$GW$. We therefore do not pay further attention to them here. 

\subsection{Local field effects in RPA}
Having understood  the band-to-band transition relation with the peaks in the optical response, we
now turn to the change in the optical response owing to the local field effects. 
First, we compare the RPA including local field effects with the long-wavelength limit in Fig. \ref{fig:rpalmf}.
In other words, this compares $\lim_{{\bf q}\rightarrow0}\varepsilon_{00}({\bf q},\omega)$ with
$\lim_{{\bf q}\rightarrow0}1/\varepsilon^{-1}_{00}({\bf q},\omega)$. 
For the in-plane polarization we can see somewhat similar peak structure but there is some tendency of shifting oscillator strength
to higher energies and there are also changes in intensity of the peaks.  For the out-of-plane polarization, we can see two
strong peaks in the RPA result around 6 and 7 eV which are not clearly related to corresponding peaks in the long-wavelength
result. These indicate that local field effects play a significant role. However, both correspond to the same onset, so the
gap does not change. 

\begin{figure}[h]
 \includegraphics[width=\linewidth]{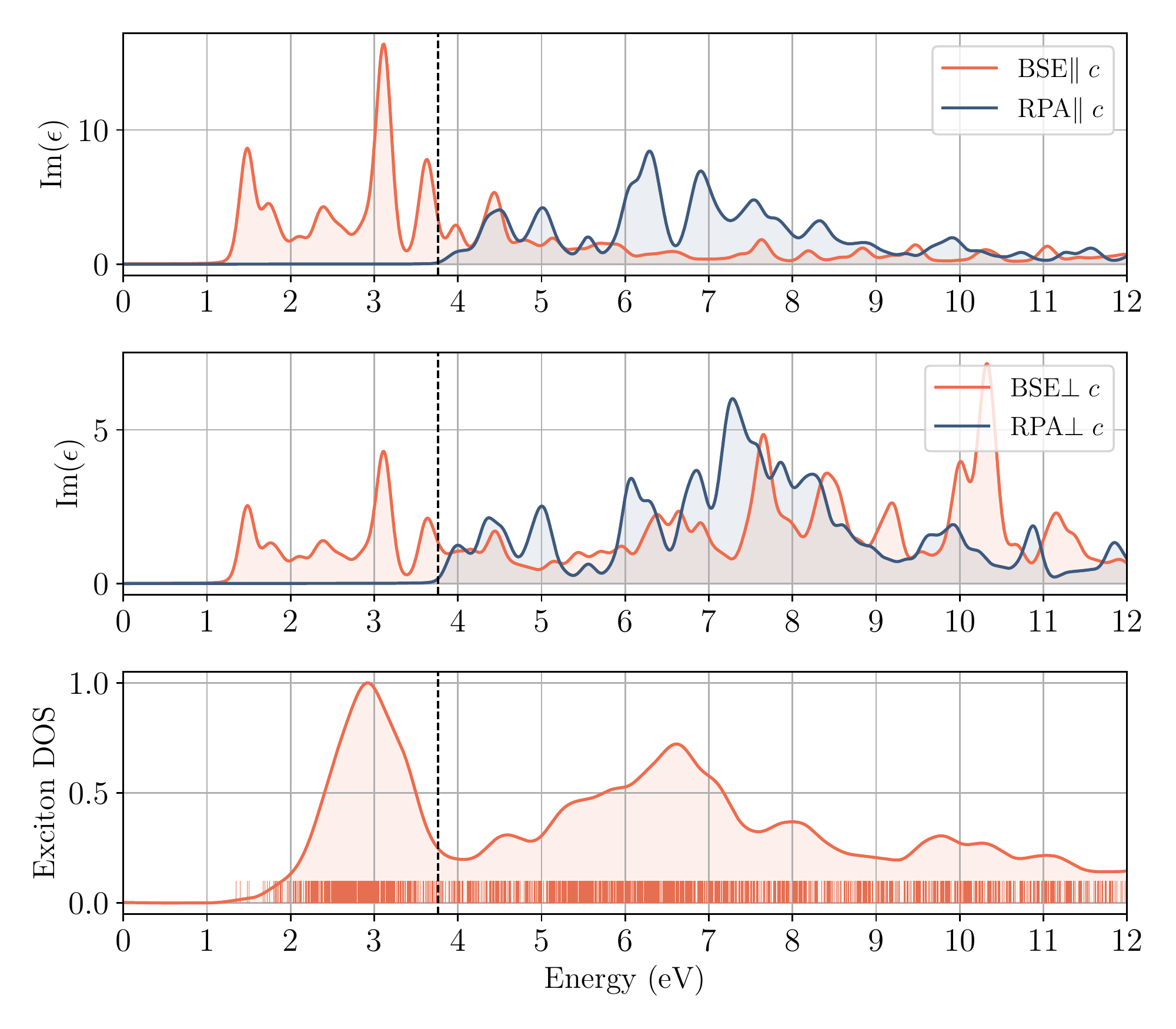}
  \caption{Macroscopic optical dielectric function $\varepsilon_2(\omega)$
    in RPA and BSE approximation both based on the QS$G\hat W$ 
    band structure. Top to bottom, $z$-polarization, in-plane polarization and exciton density of states without optical matrix elements and exciton spectrum. \label{figbserpa}}
\end{figure}

\subsection{Electron-hole effects: RPA{\sl vs.} BSE}
\begin{figure}
  \includegraphics[width=\linewidth]{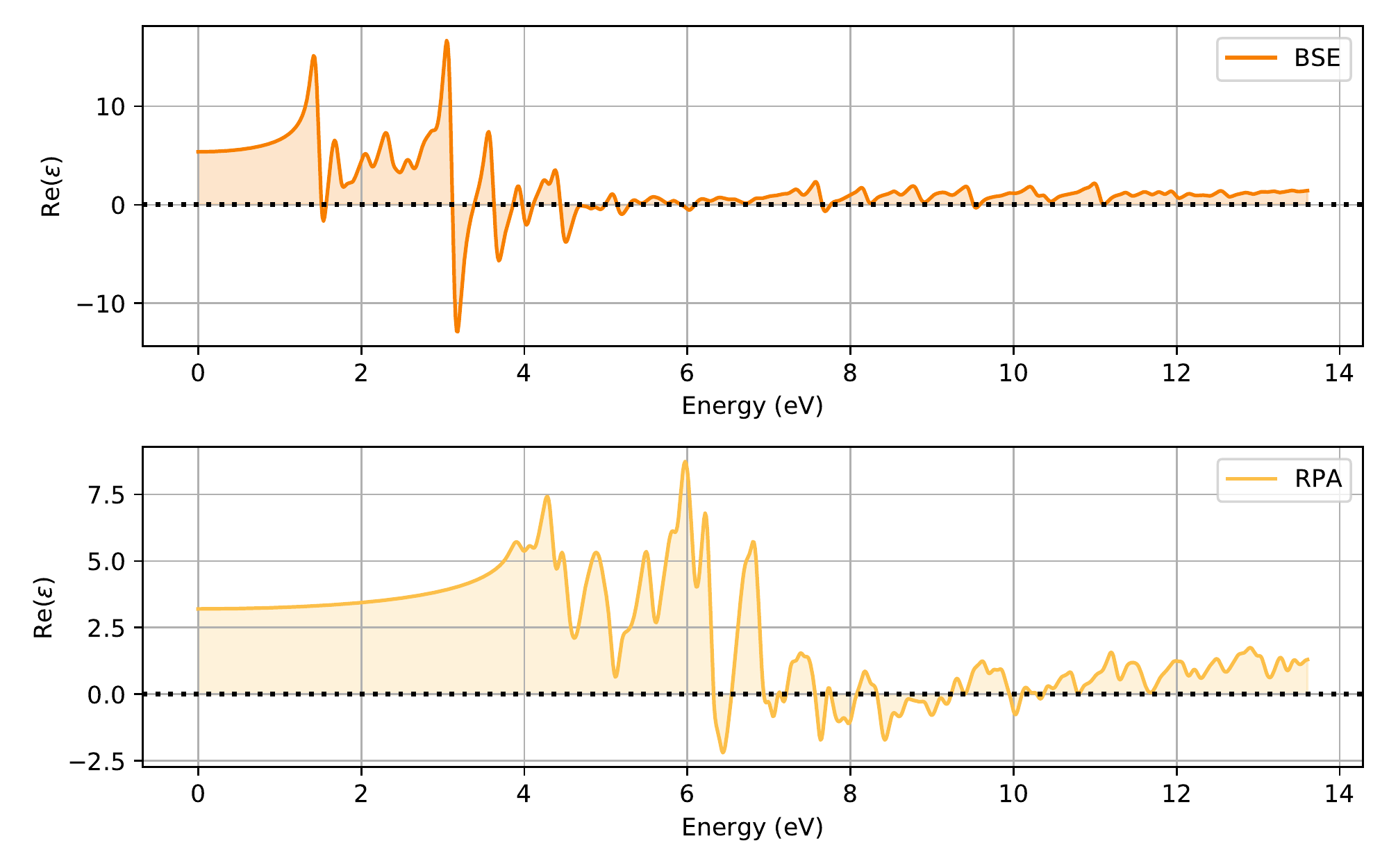}
  \caption{Real part of $\varepsilon(\omega)$ in RPA and BSE for ${\bf E}\perp {\bf c}$\label{re-eps}}
\end{figure}

Next, we show the optical dielectric functions for both polarizations
comparing the  RPA and BSE methods  in Fig. \ref{figbserpa}.
The curve labeled RPA is calculated following Eq. \ref{eqRPA}
or equivalently Eq. \ref{eqepsmac}  but using $P^0$ instead of $\bar{P}$,
(it still includes $e-h$ coupling by using $\hat W$) and the curve labeled BSE
includes electron-hole coupling via Eq.\ref{eqepsmac}.
We can see that the BSE completely changes $\varepsilon_2(\omega)$. 
There is a large shift down to lower energies. Remarkably, several peaks in the lowest energy
absorption are excitonic in nature because they lie well below the quasiparticle fundamental gap.
This indicates a very large exciton binding energy.  
This is related to the rather flat dispersion of the valence and conduction band which show
several local maxima/minima in {\bf k}-space.
It is also indicative  of the 2D character of the band structure in
this layered compound.  It is well known that lower dimensionality increases excitonic effects.
The eigenvalues of the 2-particle Hamiltonian or exciton level spectrum is shown in Fig. \ref{figldagw} in the
rightmost panel. It shows that the lowest optical gap or excitonic gap,
corresponding to the lowest exciton peak in Fig. \ref{figbserpa}, at $\sim$1.5 eV is close to the LDA gap reported in Ref. \onlinecite{Czyzyk92} 
but still somewhat larger than our GGA band structure gap.
A closer comparison with the exciton eigenvalues, shows that the lowest
exciton eigenvalue at 1.39 eV is  dark and lies below the first exciton peak in $\varepsilon_2$.  We note that using $W$ instead of $\hat W$ would increase
the quasiparticle gap but also increase the exciton binding energies and
thus the final energy of the exciton binding energy may not be
affected that much by omitting the ladder diagrams but it would
increase the difference between quasiparticle and optical gap. 

Because of this high exciton binding energy, we may consider this a Frenkel exciton and expect it also
to be quite localized in real space.  The density of exciton states
is also shown in Fig. \ref{figbserpa}. So, this shows the density of
the two particle states including the electron-hole interaction
but without weighing the intensity by optical matrix elements. 
It shows that the optical
matrix elements are quite important in determining the actual
optical spectrum. In other words, several of the two particle
eigenvalues or exciton states are dark. 
For completeness, we also show the real part of the dielectric
function in Fig.\ref{re-eps}. It shows that the downward shift of the spectrum
by BSE compared to RPA significantly affects the static
$\varepsilon_1(\omega=0)$, which is strongly enhanced.

Finally, in Appendix \ref{sm:wwhat} we show that the BSE results using
the RPA $W$ instead of the ladder $\hat W$ also shows strong excitons
with almost the same position of the lowest excition peaks. However,
in that case the difference between the fundamental gap of QS$GW$
and the exciton or optical gap is even larger because the exciton
binding energy increases with the larger $W$.

\begin{figure}
  \includegraphics[width=\linewidth]{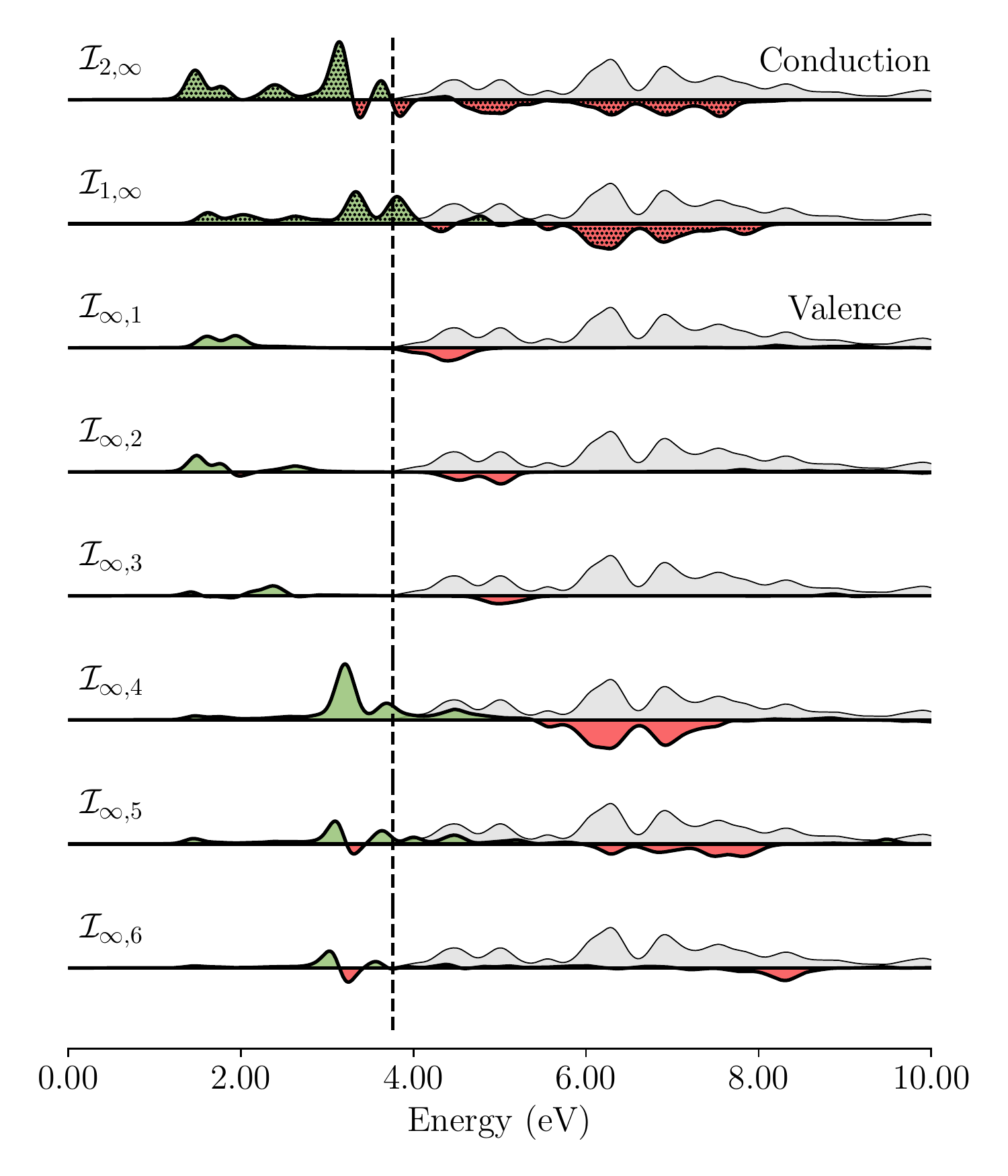}
  \caption{Contributions to the BSE optical absorption from different bands.
    The grey filled spectrum is the RPA reference spectrum.
    ${\cal I}_{1,\infty}$ shows the difference
    between BSE, while including all valence bands but only the first conduction band,  with RPA,   ${\cal I}_{2,\infty}$ shows difference between BSE
    including conduction bands 1,2 with only conduction band 1.    ${\cal I}_{\infty,1}$ means all conduction bands included but only top
    valence band in BSE relative to RPA, ${\cal I}_{\infty,2}$
    means difference between including valence bands 1,2 {\sl vs.} only 1
    while keeping all conduction bands. \label{fig:excontrib}}
\end{figure}

\subsection{Further analysis of BSE}
To gain further insight into the contributions of different bands
to the excitons, we performed separate calculations in which
we restrict the number of band pairs included
in the basis set of the BSE calculation. This analysis is shown
in Fig. \ref{fig:excontrib}. The first two sets of spectra from top to bottom
focus on the contributions of the two Co-$d$-$e_g$ conduction bands while
keeping all nine valence bands, derived from the six O-$2p$
and three Co-$d$-$t_{2g}$ bands. The thin line and grey filled curve
shows the RPA $\varepsilon_2(\omega)$ in all cases. The second spectrum
from the top shows the difference between the BSE including only
the lowest conduction band $c_1$ and the RPA,
or $\varepsilon_2^{BSE}(c_1,v_1-v_9)-\varepsilon_2^{RPA}$. In the figure we label this
as ${\cal I}_{1,\infty}$
The top curve shows the
difference due to adding the second conduction band $c_2$,
so the $\varepsilon_2^{BSE}(c_1-c_2,v_1-v_9)-\varepsilon_2^{BSE}(c_1,v_1-v_9)$.
In the figure, this is labeled as ${\cal I}_{2,\infty}$. 
In the first step, one can see that spectral weight is pulled away from the RPA 
mostly from the region above 6 eV, 
which from our previous band to band analysis corresponds to interband
transitions from the lower O-$2p$ related bands to the $e_g$ bands. 
One can see that adding the second conduction band further shifts the lowest
exciton peak down and increases its intensity, so both conduction bands
contribute to the lowest energy exciton. The excitons near 3.5 eV also
are shifted down to lower energy but stay above 3.0 eV and additional
weight is pulled from the lower peaks above the CBM to the exciton region. 

The next spectra from 3 and onward show the effect of adding valence
bands one by one while keeping both conduction bands. Thus
spectrum 3 from the top shows $\varepsilon_2^{BSE}(c_1-c_2,v_1)-\varepsilon_2^{RPA}$ (${\cal I}_{\infty,1}$)
Note that we count valence bands from top to bottom.  Spectrum
4 shows $\varepsilon_2^{BSE}(c_1-c_2,v_1-v_2)-\varepsilon_2^{BSE}(c_1-c_2,v_1)$,
and so on. These spectra show that the lowest excitons are primarily pulling
oscillator strengths from the first peaks above the gap in the RPA spectrum. 
The second valence band still shifts these down to lower energy and enhances
their intensity. But adding more valence bands to the calculation  no longer
changes this spectrum, so the change in this region goes to zero.
The exciton peak at about 2 eV comes in when we add the third valence band
and pulls spectral weight primarily from the second RPA peak above the VBM.
Finally, the peak at 3.5 eV is seen to appear only when we add valence band
$v_4$ and pulls weight from the peaks above 6 eV in RPA.
Further smaller changes occur in this exciton peaks when adding deeper
valence bands, further shifting the peak downward and pulling spectral
weight from higher and higher peaks in the RPA, which correspond to contributions
from deeper valence bands with more and more O-$2p$ character.

\begin{figure}
    \includegraphics[width=\linewidth]{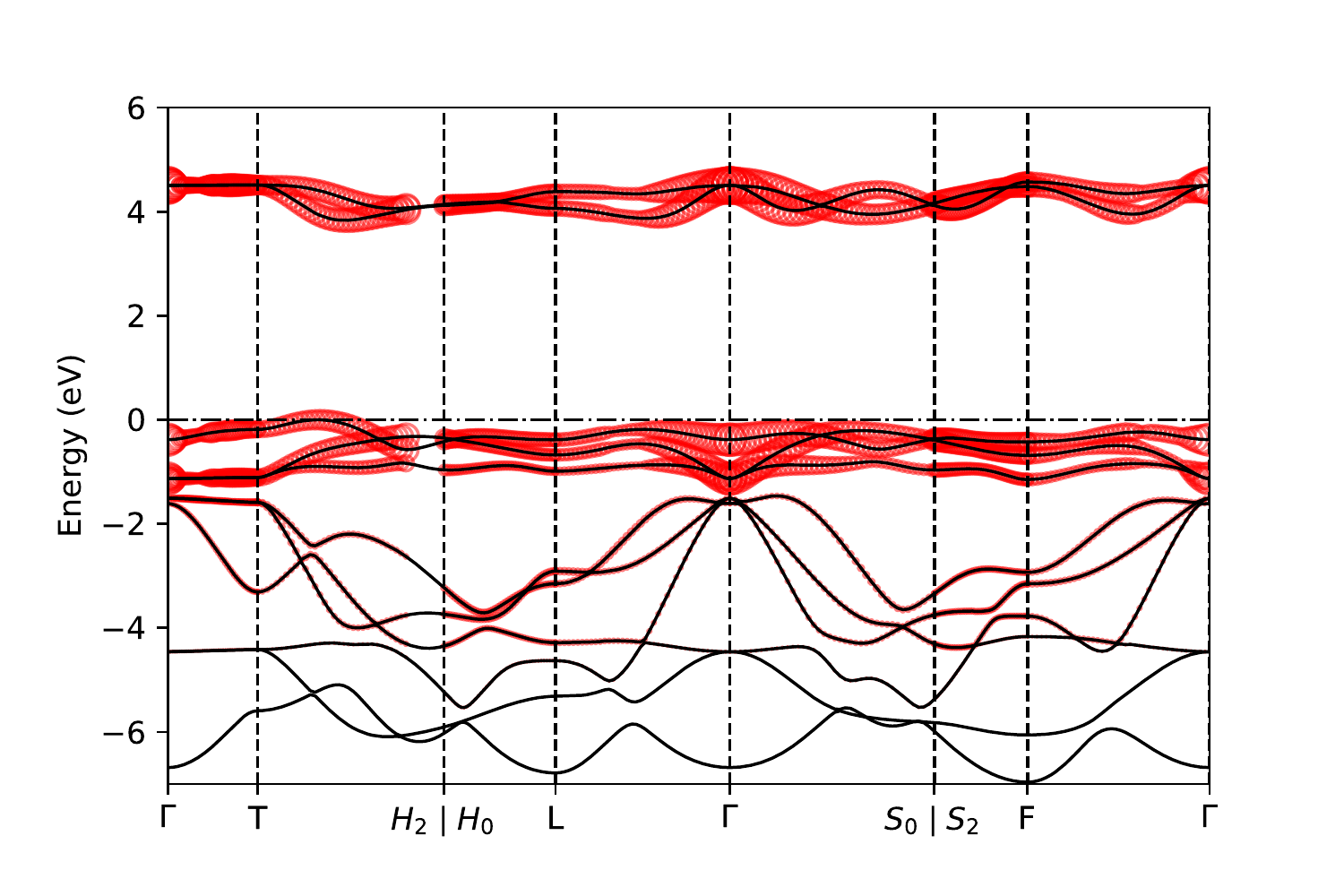}
    \caption{Band contributions to the excitons in the range 1.6-1.9 eV. \label{fig:exciton-band}}
\end{figure}
    
Next, we analyze which band pairs and {\bf k}-points contribute
to the $\varepsilon_2(\omega)$ exciton spectrum integrated over
narrow energy ranges near the peaks of the exciton spectrum.
We divide the BSE exciton region in separate intervals $[1.26-1.6]$, $[1.6-1.9]$
,$[1.9-2.21]$, $[2.21-2.76]$, corresponding each to a separate peak in the
optical spectrum. As an example, 
Fig. \ref{fig:exciton-band} shows which band state $(n,{\bf k})$ contribute to the exciton
eigenvalues in the range 1.6-1.9 eV.  We obtain this figure by using  the eigenvectors
$|A_{n,{\bf k}}^\lambda|^2$ as a weight at each {\bf k} and band $n$
where $\lambda$ indicate the exciton eigenvalue and then including all eigenvalues
in the given energy range. This is then visualized as the size of the circles on the band structure.
We can see that the top three  valence bands and lowest two conduction bands dominate the band content
of these excitons.  There is a somewhat smaller contribution from deeper valence bands.
In terms of {\bf k} distribution, the contributions are almost uniformly spread over {\bf k}-space.
Similar conclusions hold if we examine the exciton band analysis for the
other energy ranges.
This is a clear indication that the  excitons in this entire range are very broadly spread in {\bf k}-space and
consequently must be strongly localized in real space. We may thus call these excitons Frenkel excitons. 
In contrast, if we make such a plot for LiF, we see that the exciton peak is closely related with the valence band maximum and conduction band minimum near $\Gamma$ and indicates a more delocalized Wannier-type  exciton (shown in
the Appendix \ref{sm:lif}).

\subsection{Comparison with experiment} \label{sec:expt}
Unfortunately, at present no polarization dependent optical data
are available for single crystals or well oriented films. 
The excitonic part of the BSE spectrum below the gap for ${\bf E}\perp{\bf c}$ shows several peaks between 1.3 eV and 3.8 eV, which is the value of the quasiparticle gap, with the strongest peaks at
1.5 eV and 3.2 eV.  A main conclusion of our calculations is that the optical
absorption gap should be significantly smaller than the single-particle gap. 
We now attempt to compare these with the various features in
absorption reported by experimental studies.

Optical absorption involving
the Co-$d$ bands was first reported by Kushida and Kuriyama
in sol-gel grown  LiCoO$_2$ films.\cite{Kushida01,Kushida02}
They assigned an absorption peak around 2.1 eV to the transition between
filled $t_{2g}$ to empty $e_g$ bands and a higher transition around 5 eV to
Co-$t_{2g}$ to Co-$4s$, mainly based on the Aydinol \etal's \cite{Aydinol} band
structure calculations although they remark the calculations place the latter
transitions at 7 eV. They also noticed a weaker absorption feature around
0.8 eV, which they claim cannot be defect associated but is rather related
to ``electronic structure near the $e_g$ conduction band edge''. It is not
clear what that means. They noticed this broad 2.1 eV peak becomes sharper
in disordered spinel-like LiCoO$_2$ and shifts to 2.9 eV. They also compare
it with a 2.3 eV gap in rocksalt CoO and associate the shift to lower gap with
less strong covalent Co-$d$-$e_g$ to O-$2p$ interaction due to the larger
Co-O distance. 

Optical absorption of Li$_x$CoO$_2$ as
function of $x$ was reported by Liu \etal,\cite{Liu15}  who associate the
1.7 eV peak which they find to be present  only in Li$_x$CoO$_2$ with $x<0.72$ to
transitions from a peak in the filled $t_{2g}$ to empty $t_{2g}$ states
within the same band and a peak at 3.08 eV to $t_{2g}$-$e_g$
transitions and higher peaks around
4.55 and 5.76 eV to transitions from  lower more O-$2p$ hybridized valence bands to the $e_g$ conduction band.

Ghosh \etal \cite{Ghosh07} determine  the gap at about 1.43 eV by differentiating  the absorption spectrum. They claim this is in good agreement with the
earlier band calculations by Czy\'{z}yk \etal \cite{Czyzyk92} although
that paper mentions a calculated gap of only 1.2 eV.
Instead, we identify this feature and the closely associated peak at 1.7 eV
with our lowest excitonic peak at 1.5 eV. The 3.0 eV feature
seen by Liu \etal \cite{Liu15} could be compared with our exciton peak
at 3.2 eV.
Lower energy features at 0.88 and 0.65 eV
in Ghosh \etal were there associated with defects but are similar to those observed by Kushida \etal in other phases
of LiCoO$_2$ only.
On the other hand, our analysis of the band-to-band origin of peaks in absorption shows that transitions from the highest valence band to higher conduction bands, beyond the $e_g$ Co bands, related to Co-$4s$ or Li occur at much
higher energy and features near 5 eV in optical absorption must rather be due
to transitions from deeper valence bands to the same Co-$e_g$ conduction band.

Balakrishnan \etal\cite{Balakrishnan}
used Tauc plots to extrapolate a gap between 2.5 and 3.24 eV depending on
annealing temperature of their sol-gel grown films. The largest gap value here
corresponded to the highest annealing temperature. These data also show a lower peak in optical absorption between 2.5 and 3.0 eV.
Rao \cite{Rao2010} also used Tauc plots  (on pulsed Laser deposited films)
but places the onset of absorption at about 2.3-2.4 eV.
The Tauc method consists in plotting 
the absorption coefficient squared and then extrapolating a linear slope
fitted to the low energy region.  However, we caution that the Tauc
plot analysis to determine the gap is based on
the expectation of a square-root behavior of the joint density of
states just above the threshold and is expected to be
valid only for direct gap semiconductors in the absence of excitons, for example
at high enough temperature that the excitons become dissociated. 
Our calculations show that the lowest peaks in optical absorption
are excitonic in nature with high exciton binding energies and
do not support the validity of a Tauc model analysis. 

As for the various experimental observations as function of temperature, 
we note that at higher
temperature, one may expect interdiffusion of Li inside the CoO$_2$ layer and formation of disordered rocksalt or spinel type
phases.\cite{Volkova21}  These could have significantly different optical
absorption properties because they have a different electronic structure.
These structures would lack the 2D like strong binding energy and
would be characterized by disorder in the Li locations which may also
have strongly correlated aspects by the deviations from the ideal
octahedral splitting of the Co $d$ bands into filled $t_{2g}$ and empty
$e_g$ bands.

Support for our assertion that the quasiparticle gap is significantly larger
than the optical absorption gap is provided by the
Bremsstrahlung isochromat spectroscopy (BIS) or inverse photoemission
and X-ray photoemission spectroscopy (XPS) by van Elp \etal\cite{vanElp}. Although their theoretical
analysis in terms of a cluster model places
the experimental gap at 2.7$\pm0.3$ eV, the BIS-XPS actual data in their figures
are consistent with a quasiparticle gap larger than 3 eV.

An overview of the different experimental and theoretical band gap and
associated optical features is given in Table \ref{taboverview}.

\begin{table}
  \caption{Overview of band gap related optical features in theory and experiment.\label{taboverview}}
  \begin{ruledtabular}
    \begin{tabular}{llll}
      &$E_g$ (eV) & ref. & comment \\ \hline
      Theory & 1.2 & \cite{Czyzyk92} & LDA \\
      & 0.867 & this work & GGA \\
      & 4.125 &    ``       & $E_g^{qp}$ QS$GW$ \\
      & 3.762 &    ``       & $E_g^{qp}$ QS$G\hat W$ \\
      & 1.5   &     ``      & $E_g^{opt}$ lowest bright exciton \\
      &&&      peak \\ \hline
      Experiment & 2.1   & \cite{Kushida01} & absorption peak $t_{2g}-e_g$ \\
      & 1.43  & \cite{Ghosh07} & differentiation of absorption \\
      & 2.5   & \cite{Balakrishnan} & Tauc plot \\
      & 2.3   & \cite{Rao2010} & Tauc plot \\
      & 1.7   & \cite{Liu15} & optical absorption \\
      & $>$3.0 & \cite{vanElp} & XPS/BIS \\
    \end{tabular}
  \end{ruledtabular}
\end{table}

Thus, our lowest exciton peak at 1.5 eV and an additional peak
at about  3.2 eV may be considered to be consistent
with experimental observations of the onset of optical absorption. 
However, the sharp nature of the excitonic peaks predicted by the BSE
is apparently not seen in the experiment. This may be because of the finite
temperature broadening, electron-phonon coupling and so on, which
are not yet taken into account by our present calculation, which only used
an {\sl ad-hoc} broadening parameter in the BSE spectra. 
We stress that the present BSE calculations provide a radically different
interpretation of the optical absorption onset than early LDA band structure
interpretations. 

Clearly, additional experimental work on single crystal samples including polarization dependence  would be useful to further explore the nature of the optical absorption onset and confirm our conclusion of its 
excitonic nature.

\section{Conclusions}
The main conclusion of this work is that the onset of optical absorption
in LiCoO$_2$ is excitonic in nature and shows a significant binding energy 
with respect to the independent particle band gap as obtained in
a $GW$ approximation and which would correspond to the difference between
BIS and XPS experiments.  More precisely we find that the QS$GW$ gaps  of 4.15 eV
is much larger than the GGA gap of 0.87 eV. The  electron hole interactions increase the
screening of $W$ and lower the QS$GW$ gap by about 11 \% but the quasiparticle gap is then still at
about 3.76 eV and cannot explain the lower onsets of absorption. However, the excitonic BSE spectrum has an onset which
is shifted back down well below the quasiparticle gap and shows a lowest excitonic peak near 1.5 eV  which agrees
with the experimentally found absorption onset at 1.4 eV
and lowest absorption peaks  near 2 eV.  A second excitonic
peak at about 3.3 eV is also hinted at in some experimental results. 
Nonetheless significant questions about the detailed interpretation of the experiments remain because of possible
complications with disordered phases of LiCoO$_2$, variation in Li content,  and so on.

\acknowledgements{The work at CWRU (SKR and WL) was supported by the Air Force Office of Scientific Research under grant number
  FA9550- 18-1-0030. The calculations made use of the High Performance Computing Resource in the Core Facility for Advanced Research Computing at Case Western Reserve University. BC, MG, DP and MvS  are grateful for support from the Engineering and Physical Sciences Research Council (EPSRC), under grant EP/M011631/1. MvS and DP are supported by the National Renewable Energy Laboratory.}

\appendix
\section{Convergence}\label{sm:convergence}
\begin{figure}[h]
  \includegraphics[width=\linewidth]{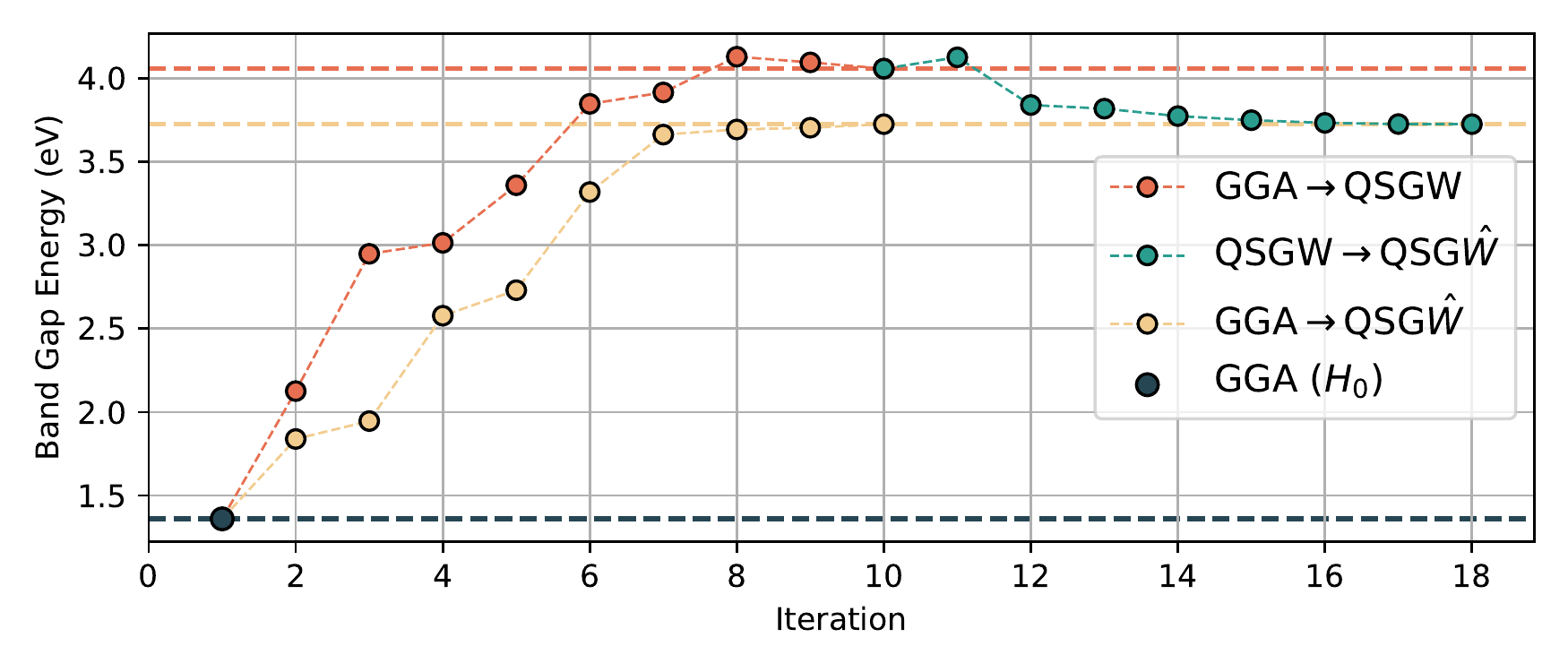}
  \caption{Convergence of the gap as function of iteration number in
    QS$GW$ and QS$G\hat W$.\label{figconv}}
\end{figure}

Fig. \ref{figconv} shows the convergence of the gap in QS$GW$ followed
by several iterations in which ladder diagrams are included using $\hat W$
compared with using the ladder diagrams from the start.
\section{BSE with $W$ vs. $\hat W$} \label{sm:wwhat}
\begin{figure}
\includegraphics[width=\linewidth]{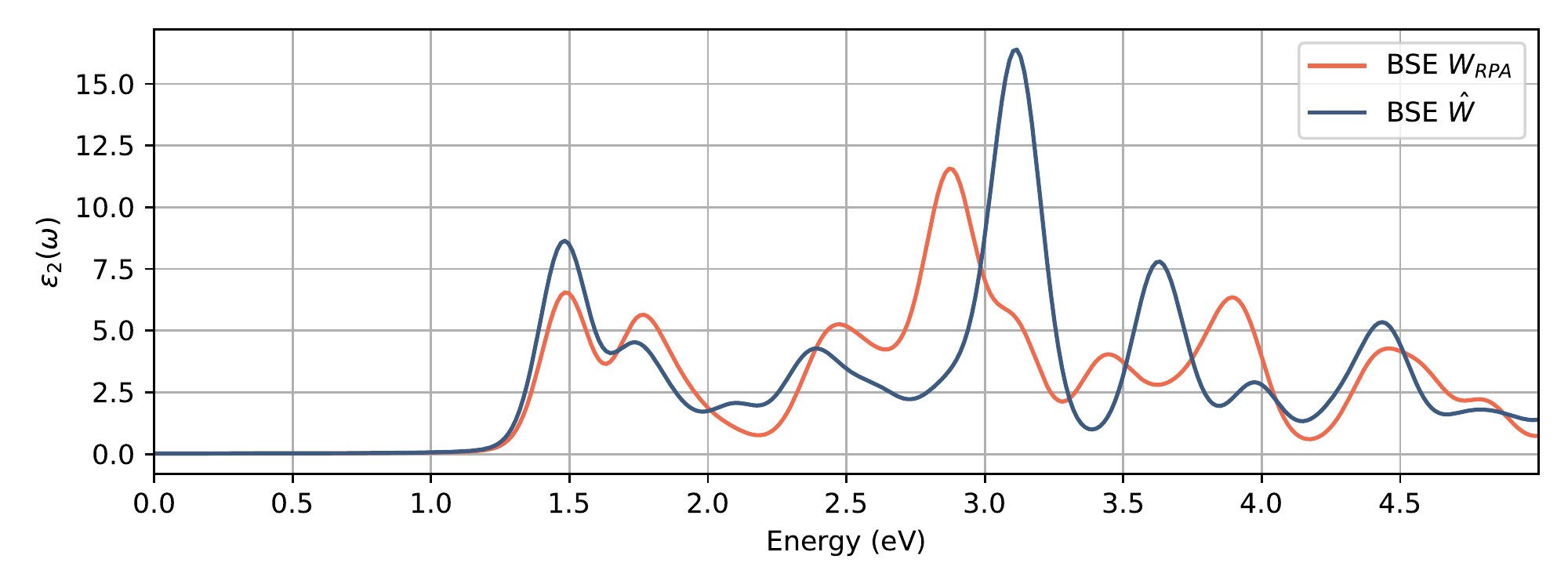}
\caption{Comparison of imaginary part of the dielectric function
  $\varepsilon_2(\omega)$  (for ${\bf E}\parallel {\bf c}$)
  with $W$ calculated in RPA and $\hat W$
  calculated including ladder diagrams both in the underlying
  band structure and BSE. \label{figbsewwhat}}
\end{figure}
Fig. \ref{figbsewwhat} shows that $\varepsilon_2(\omega)$ calculated with
$W$ in the RPA is similar but not identical to the one calculated with $\hat W$
including ladder diagrams. This is because the larger $W$ in the RPA increases
the fundamental gap of QS$GW$ but also increases the exciton binding energies
in a proportional manner. 
\section{LiF}\label{sm:lif}
\begin{figure}[h]
    \includegraphics[width=\linewidth]{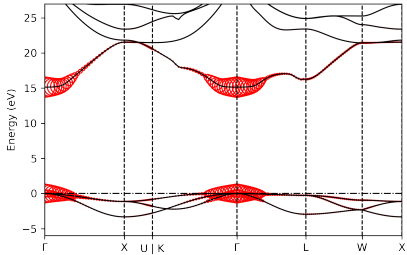}
    \caption{Exciton projected band structure for LiF. \label{fig:exciton-lif}}
\end{figure}

To distinguish the Frenkel excitons found in this paper for LiCoO$_2$
which have a very delocalized origin in band states in {\bf k}-space,
we show here  in Fig. \ref{fig:exciton-lif} the corresponding figure for
LiF. The good agreement with other calculations,
for example using the exciting code, \cite{exciting,Vorwerk19,nomad}
for this {\bf k}-space origin
of the exciton in LiF, which has Wannier exciton character
also serves to demonstrate
the validity of our methodology. 

\bibliography{Bib/lmto,Bib/dft,Bib/gw,Bib/licoo2,Bib/bse}
\end{document}